\documentclass[conference]{IEEEtran}
\usepackage{cite}
\usepackage{amsmath,amssymb,amsfonts}
\usepackage{algorithmic}
\usepackage{graphicx}
\usepackage{textcomp}
\usepackage{xcolor}
\usepackage[hidelinks]{hyperref}
\usepackage{url}
\usepackage{placeins}
\usepackage{comment}
\DeclareUnicodeCharacter{03A8}{$\Psi$}
\usepackage{booktabs}

\makeatletter
\newcommand{\linebreakand}{%
  \end{@IEEEauthorhalign}
  \hfill\mbox{}\par
  \mbox{}\hfill\begin{@IEEEauthorhalign}
}
\makeatother
\def\BibTeX{{\rm B\kern-.05em{\sc i\kern-.025em b}\kern-.08em
    T\kern-.1667em\lower.7ex\hbox{E}\kern-.125emX}}
\begin{document}

\title{Toward Accessible Psychotherapy Training Using AI-Driven Interactive Patient Avatars\\}

\author{\IEEEauthorblockN{Pascal Riachi}
\IEEEauthorblockA{
\textit{ETH Zurich}\\
Switzerland \\
priachi@ethz.ch}
\and
\IEEEauthorblockN{Sofie Kamber}
\IEEEauthorblockA{
\textit{ETH Zurich}\\
Switzerland \\
skamber@ethz.ch}
\and
\IEEEauthorblockN{Stella Brogna}
\IEEEauthorblockA{
\textit{University of Lucerne}\\
Switzerland \\
stella.brogna@unilu.ch}
\linebreakand
\IEEEauthorblockN{Andrew Gloster}
\IEEEauthorblockA{
\textit{University of Lucerne}\\
Switzerland \\
andrew.gloster@unilu.ch}
\and
\IEEEauthorblockN{Rafael Wampfler}
\IEEEauthorblockA{
\textit{ETH Zurich}\\
Switzerland \\
rafael.wampfler@inf.ethz.ch}
}

\maketitle

\begin{abstract}
Training psychotherapists in evidence-based interventions such as Acceptance and Commitment Therapy (ACT) requires repeated practice with meaningful feedback, yet opportunities for safe, standardized training are limited by ethical, logistical, and resource constraints. We introduce a system designed to support ACT-oriented psychotherapy training through spoken dialogue with an embodied virtual patient. The system uses large language models to simulate patient behavior conditioned on profiles derived from real therapy sessions and configurable clinical scenarios, while a separate automated evaluator provides turn-by-turn feedback on therapist responses based on established ACT fidelity criteria. Rather than aiming to replace supervision, the system is intended to support deliberate practice by enabling experimentation, reflection, and immediate feedback in low-risk settings. Expert evaluation with practicing psychologists confirmed high realism in patient behavior and demonstrated that immediate turn-by-turn ACT feedback increased therapists' awareness of intervention choices and enabled effective experimentation with alternative responses. Quantitative evaluation across 49 therapy transcripts identified GPT-4o-mini as the optimal feedback model, achieving the lowest mean absolute error (MAE = 6.12) in replicating human supervisor ACT fidelity ratings with statistically significant agreement. This work demonstrates the potential of fidelity-aware simulated patients as a scalable complement to psychotherapy training.
\end{abstract}



\section{Introduction}

Training clinicians to deliver evidence-based psychotherapy interventions competently remains challenging. Acceptance and Commitment Therapy (ACT), an evidence-based behavioural intervention that aims to increase psychological flexibility through processes such as acceptance, cognitive defusion, values clarification, and committed action \cite{hayes_acceptance_2006,hayes_acceptance_2012}, exemplifies this challenge. Traditional training approaches often improve theoretical knowledge but do not consistently translate into changes in therapist behaviour \cite{frank_therapist_2020}. Psychotherapy supervision alone yields mixed effects and is constrained by variability in quality, access, and resources \cite{vaz_rethinking_2025}. While standardised role-plays and simulated patient interactions show promise, their use in psychotherapy training remains limited \cite{kuhne_standardized_2020}. The persistent gap between knowledge and competence highlights the need for deliberate practice --- repeated, goal-directed exercises with immediate feedback in controlled environments \cite{berning_effects_2024}. However, opportunities for such practice are constrained by ethical considerations, limited availability of standardised patients, and resource-intensive supervision \cite{blackmore_simulation-based_2018}.
Recent advances in large language models (LLMs) and conversational AI present new opportunities to create interactive training systems that simulate patient behaviour while providing structured feedback aligned with therapeutic fidelity measures, offering unlimited practice in safe, controlled settings \cite{blanco_integrating_nodate}. 

We present a simulated patient training application combining an embodied avatar with LLMs to enable interactive ACT-oriented psychotherapy practice. Virtual patient behaviour is conditioned on profiles derived from real therapy transcripts, enabling clinically relevant interactions. The system provides realistic patient simulation responding dynamically to therapist interventions and automated turn-by-turn feedback evaluating responses against ACT Fidelity Measure (ACT-FM) dimensions \cite{goodyer_treatment_2017}.

Expert evaluation with two practising psychologists confirmed that the system produces realistic patient interactions and that immediate ACT-aligned feedback effectively increases therapeutic awareness and supports deliberate practice. Through quantitative evaluation across 49 therapy sessions, we identified GPT-4o mini as achieving closest alignment with human ACT fidelity ratings (MAE = 6.12, $p < 0.001$), establishing a validated approach for automated real-time feedback. These findings demonstrate the feasibility of scalable, fidelity-aware AI-supported psychotherapy training.

\subsection{Contributions}
The contributions of this paper are threefold: 
\begin{itemize}
    \item An embodied simulated patient environment enabling spoken dialogue practice with clinically grounded virtual patients across configurable ACT training scenarios.
    \item An automated ACT fidelity scoring mechanism providing immediate turn-by-turn feedback aligned with ACT-FM dimensions during live interaction.
    \item A systematic evaluation of six LLMs for ACT fidelity assessment, identifying GPT-4o mini as optimal based on agreement with human expert ratings.
\end{itemize}
\section{Related Work}

\subsection{Virtual Patients for Clinical Training}
Virtual patients have been widely studied as an interactive simulation modality for health professions education, offering learners opportunities to engage with realistic clinical scenarios in a safe environment where clinical decision-making and communication skills can be practised without risk to real patients \cite{kononowicz_virtual_2019}. Systematic reviews suggest that virtual patients can support clinical reasoning and related competencies through diverse case presentations and reflective practice \cite{kononowicz_virtual_2019}. Emerging evidence indicates that virtual simulation tools can enhance interpersonal competencies in clinical trainees \cite{fernandez-alcantara_virtual_2025}. Prior work has explored embodied virtual patients through VR and virtual human technologies to enhance immersion and realism across medical interviewing, emergency response \cite{stansfield_design_2000}, and procedural training \cite{talbot_virtual_2019}.

\subsection{Conversational AI for Psychotherapy Training}

Recent advances in conversational AI and large language models have enabled interactive systems that engage users in psychologically meaningful dialogue, motivating their application within psychotherapy training \cite{naswa_assessing_2024}. Virtual patient systems have focused on simulating client behavior to enable structured practice. Patient-Ψ conditions LLMs on structured psychological profiles to generate psychotherapy-oriented dialogue, emphasizing explicit modeling of patient characteristics for behavioral consistency \cite{wang_patient-_2024}. Petrizzo-VP demonstrates that conditioning language models on therapy transcript excerpts yields more human-like patient responses than abstract behavioral descriptions \cite{petrizzo_building_2025}. Embodied virtual patients have also been explored for communication skills training in psychiatric and geriatric care \cite{chaby_embodied_2022}. Unlike prior systems that focus primarily on patient simulation, our work integrates real-time, turn-by-turn automated feedback based on ACT-FM dimensions and systematically evaluates multiple LLMs to identify the optimal model for fidelity-aligned feedback generation.

\subsection{ACT Fidelity, Empathy, and Automated Assessment}
\label{rel-work:act}
Fidelity assessment evaluates adherence to treatment model principles \cite{carroll_general_2000}, but reliance on expert human raters makes it time-consuming and hard to scale \cite{perepletchikova_treatment_2005}. ACT targets six core processes --- acceptance, cognitive defusion, present-moment contact, self-as-context, values clarification, and committed action --- which therapists are expected to enact experientially rather than didactically \cite{hayes_acceptance_2006, hayes_acceptance_2012}. The ACT Fidelity Measure (ACT-FM) operationalizes adherence through 25 items across four paired dimensions capturing ACT-consistent behaviors (e.g., acceptance, cognitive defusion, values) alongside ACT-inconsistent ones (e.g., lecturing, control-focused language, cognitive fusion), rated 0--3 \cite{oneill_development_2019}. An ACT balance score is computed as ACT-consistent minus ACT-inconsistent totals. The Therapy Empathy Scale (TES) complements this with nine empathy facets (e.g., warmth, emotional resonance, acceptance) rated 1--7 and averaged into an overall score \cite{decker_development_2014}. Recent work shows LLMs can approximate human fidelity and empathy ratings in therapeutic dialogue \cite{tahir_thinking_2025}. We extend this by implementing turn-by-turn real-time feedback and systematically comparing six LLMs for automated ACT fidelity assessment.

\section{Method}

\subsection{System Overview}

\begin{figure*}
    \centering
     \includegraphics[width=\textwidth, trim=0.0cm 1.7cm 0cm 1cm, clip]{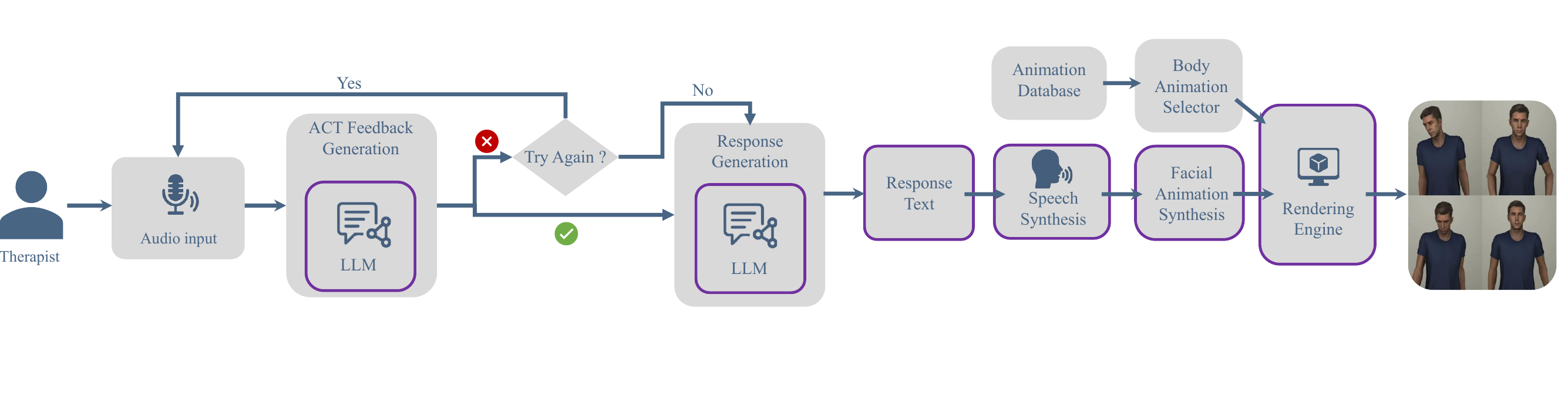}
    \caption{System architecture of the psychotherapy training application. Therapist utterances are transcribed from audio and evaluated for ACT-consistency by a feedback LLM. A separate response LLM produces patient responses conditioned on patient profiles, which are rendered through synthesized speech, facial animation, and body animations.}
    \label{fig:system_architecture}
\end{figure*}

We developed a simulated patient training application in Unity (Figure~\ref{fig:system_architecture}) that enables interactive psychotherapy practice through real-time spoken dialogue using Azure Cognitive Services for speech processing. The system employs a dual-LLM architecture: GPT-4o generates contextually appropriate patient responses based on structured prompts (Table~\ref{tab:vp-response-prompt}) combining patient profiles and training scenarios, while GPT-4o mini serves as an automated evaluator that assesses each therapist utterance for ACT-consistency and provides immediate feedback. Therapist responses are captured via voice input, transcribed, and evaluated by the feedback LLM. The patient simulation LLM then generates a response rendered through Azure text-to-speech synthesis and synchronized with embodied avatar animations. The application is designed exclusively for training and research purposes, enabling therapists to practice clinical skills in a controlled simulation environment without the ethical, safety, or logistical risks of real patient encounters.

\begin{table}[t]
\centering
\caption{Structure of the Response-Generation Prompt Constructed from a Patient Profile and an Active Training Scenario.}
\label{tab:vp-response-prompt}
\begin{tabular}{p{0.32\linewidth} p{0.6\linewidth}}
\hline
\textbf{Component} & \textbf{Description} \\
\hline
Role Instruction 
& Defines the virtual patient identity and constrains responses to the patient perspective only. \\

Patient Profile 
& Transcript-derived background including demographics, core difficulties, emotional and cognitive themes, values, behavioral patterns, and current situation. \\

Active Scenario 
& Scenario-specific modifiers (e.g., suicidality, resistance) that adjust affective tone, conversational stance, emotional intensity, and response emphasis for the current session. \\

Response Style Guidelines 
& Constraints on turn length, linguistic patterns, social behavior (e.g., no therapist-directed questions), and conversational rhythm to ensure authentic patient behavior. \\

Safety Boundaries
& Scope and limits for discussing sensitive content to ensure clinically appropriate and ethically responsible behavior. \\

Illustrative Excerpts 
& Selected anonymized dialogue excerpts from real transcripts grounding responses in realistic language and interaction patterns. \\
\hline
\end{tabular}
\end{table}

\subsection{Patient Profiles}

Virtual patient behavior is conditioned on four anonymized patient profiles (two male, two female), each constructed from 4--7 real ACT psychotherapy sessions from \cite{gloster_psychotherapy_nodate}. Sessions averaged 58--92 therapist-patient turn pairs and lasted 38--60 minutes. Transcripts were analyzed jointly to capture stable patterns of experience, behavior, and interaction that persist across sessions.

Patient profiles were synthesized using GPT-5.2 via Azure with a structured prompt that instructed the model to analyze multiple sessions jointly and extract stable patterns across seven dimensions (Table~\ref{tab:patient-profile}): patient overview, core difficulties, values and motivations, emotional and cognitive themes, behavioral patterns, progress trajectory, and simulation hooks. The prompt explicitly constrained the model to rely only on transcript content, avoid unsupported details, produce anonymized descriptions, and extract representative dialogue excerpts illustrating emotional tone and interaction style. This approach follows prior findings that transcript-conditioned representations yield more coherent and human-like virtual patient behavior than abstract behavioral descriptions~\cite{petrizzo_building_2025}. These profiles (Table~\ref{tab:patient-profile}) are incorporated into the LLM prompt as persistent contextual information to enable coherent, clinically plausible behavior throughout a session.

\begin{table}[ht]
\centering
\caption{Structure of a Transcript-Derived Virtual Patient Profile Used to Condition Patient Behavior During Simulation.}
\label{tab:patient-profile}
\begin{tabular}{p{0.3\linewidth} p{0.6\linewidth}}
\hline
\textbf{Profile Section} & \textbf{Content} \\
\hline
Patient Overview 
& Anonymized demographic approximation and life context relevant to therapy, expressed in neutral descriptive language. \\

Core Difficulties 
& Summary of primary psychological challenges and recurrent concerns across sessions. \\

Values and Motivations 
& Salient values and motivational orientations inferred from patient statements and therapeutic goals. \\

Emotional and Cognitive Themes 
& Common emotional states and characteristic thought patterns shaping the patient’s experience. \\

Behavioral Patterns 
& Typical coping strategies, avoidance behaviors, and interpersonal styles exhibited in therapy. \\

Progress Trajectory 
& Overview of changes, insights, and fluctuations observed over the course of sessions. \\

Simulation Hooks 
& Key triggers, affective tone, and clinically relevant entry points designed to support scenario-based training. \\

Representative Excerpts 
& Selected anonymized dialogue snippets illustrating emotional tone and interaction style. \\
\hline
\end{tabular}
\end{table}

\subsection{Training Scenarios and Exercises}

The system allows therapists to select from predefined training scenarios (suicidality, resistance, lack of motivation, heightened anxiety, therapeutic rupture, homework non-compliance, dramatic escalation) that modulate virtual patient behavior during interaction. Each scenario is combined with a compatible patient profile, introducing situational modifiers that adjust affective tone, conversational stance, and response priorities without overriding core profile characteristics. Scenario–profile pairings were manually reviewed to ensure clinical plausibility. For example, suicidal ideation scenarios are appropriate for profiles characterized by burnout and shame but not for profiles centered on interpersonal conflict without depressive features.

\subsection{Automated Feedback Generation}

The system integrates an automated feedback component to support therapist learning during interaction. GPT-4o mini via Azure evaluates therapist responses turn-by-turn following speech transcription, starting from the third therapist turn to ensure sufficient conversational context.

The feedback LLM is guided by a structured prompt that frames the model as an ACT clinical supervisor with explicit instructions to: (1) prioritize the most recent therapist utterance as primary evidence for scoring, treating earlier turns only as interpretive context; (2) score based on intervention function (openness vs. control) rather than form; (3) reward experiential over conceptual work; and (4) evaluate workability rather than thought accuracy. At each turn, input consists of the full dialogue history with the current therapist utterance explicitly marked as the evaluation target.

The model produces structured JSON output containing scores for all 25 ACT-FM items rated on a 0--3 Likert scale (0 = never, 3 = consistently), organized across four paired dimensions: therapist stance, open response style, aware response style, and engaged response style. Each dimension captures both ACT-consistent behaviors (e.g., experiential methods, cognitive defusion, present-moment focus, values clarification) and ACT-inconsistent behaviors (e.g., lecturing, control-focused interventions, formulaic mindfulness, imposed values). When any ACT-inconsistent items are scored above zero, the output includes a single-sentence rationale explaining why the final utterance reflects ACT-inconsistent behavior.

Individual item scores are aggregated into an \emph{ACT balance} score. Because ACT-FM dimensions contain different numbers of items, ACT-consistent stance items (items 1--4) are averaged and scaled to match the three-item structure of other dimensions, then summed with remaining ACT-consistent scores and contrasted against summed ACT-inconsistent scores. Non-negative ACT balance indicates ACT-consistent responses; negative balance indicates ACT-inconsistent responses.

\begin{figure}
    \centering
    \begin{minipage}[t]{0.28\linewidth}
        \centering
        \includegraphics[width=\linewidth]{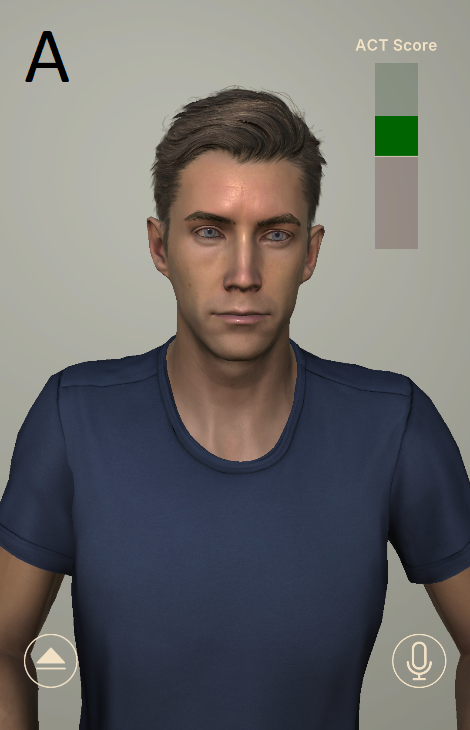}
    \end{minipage}
    \hspace{0.01\linewidth}
    \begin{minipage}[t]{0.28\linewidth}
        \centering
        \includegraphics[width=\linewidth]{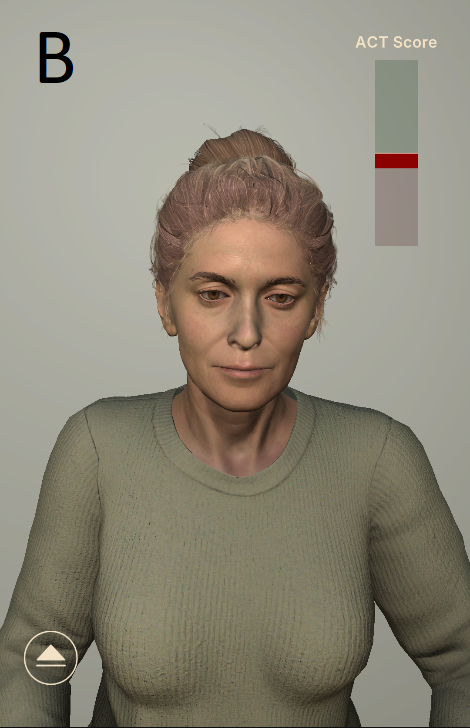}
    \end{minipage}
    \hspace{0.01\linewidth}
    \begin{minipage}[t]{0.28\linewidth}
        \centering
        \includegraphics[width=\linewidth]{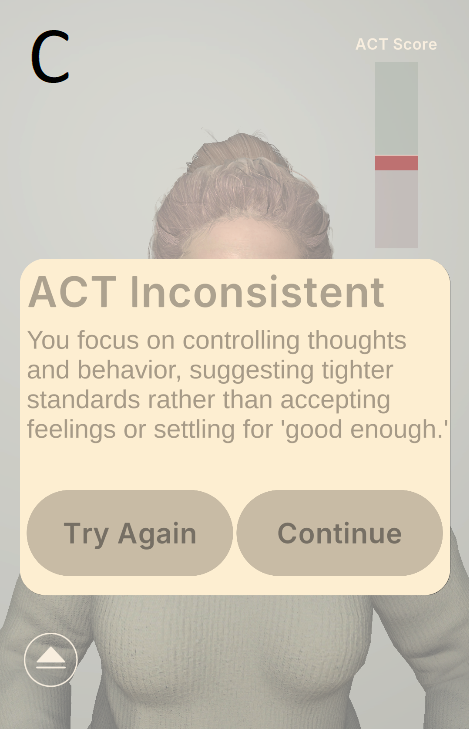}
    \end{minipage}
    \caption{Automated ACT-aligned feedback interface. (a) ACT-consistent feedback displayed as a green balance indicator. (b) ACT-inconsistent feedback displayed as a red balance indicator. (c) Modal dialog providing rationale and options to retry or continue when responses are ACT-inconsistent.}
    \label{fig:act_feedback}
\end{figure}

Feedback is presented through a visual ACT balance indicator (Figure~\ref{fig:act_feedback}). Positive values display as a proportionally filled green bar; negative values display as a downward-filling red bar. When ACT balance is negative, a modal dialog pauses the session, presents the evaluator's rationale, and prompts the therapist to retry or continue. Selecting \emph{retry} removes the most recent utterance, allowing reformulation and immediate observation of alternative interventions' effects on the ACT score. Selecting \emph{continue} registers the response and resumes dialogue, supporting iterative skill refinement.

\subsection{Avatar Animation and Embodiment}

The virtual patient is presented through photorealistic embodied avatars (one male, one female) rendered in Unity using Character Creator 4~\cite{noauthor_character_nodate}. Higher visual realism can enhance user engagement, social presence, and perceived credibility in training contexts~\cite{montanha_crafting_2023}. Patient responses are converted to speech using Azure text-to-speech and synchronized with facial and body animations. Facial animation uses NVIDIA Audio2Face~\cite{nvidia_audio2face-3d_2025}, which maps speech audio to detailed facial movements. Eight body gestures were selected from 23 candidates by two expert psychotherapists, rated for clinical plausibility, emotional congruence, and affective range; randomized cycling prevents mechanical repetition. Subtle micro-animations (blinking, head movement, gaze) added using SALSA Emoter~\cite{salsa2016lip} were prioritized because non-verbal cues are key predictors of perceived agent naturalness and trainee engagement~\cite{montanha_crafting_2023}.
\section{Evaluation and Results}
\subsection{Expert Evaluation}

To assess system usability, realism, and training relevance, we conducted a mixed-methods evaluation with two female psychologists (ages 25--30) who completed simulated ACT-oriented sessions. One was a licensed psychotherapist with 4--7 years of clinical experience and high ACT familiarity; the other was a psychotherapist in training with 1--3 years of experience and moderate ACT familiarity. Both had prior role-play training experience. Each participant interacted with the system for approximately 90 minutes across multiple scenarios including suicidality, resistance, and dramatic escalation.

Both psychologists rated the virtual patient as realistic, with natural language, emotional expression, and appropriate responsiveness to interventions. Patient reactions were coherent with active scenarios and introduced clinically meaningful challenges. Participants reported temporarily suspending disbelief and engaging comparably to role-play, though they noted the virtual patient resembled an ``ideal'' training case --- someone with prior therapy experience who engages reflectively with ACT interventions.

Both participants valued the immediate, turn-by-turn ACT feedback, reporting that the visual ACT score and retry/continue mechanism increased awareness of therapeutic language and intervention choices. The retry option enabled experimentation without disrupting dialogue flow.

Participants identified limitations: the virtual patient sometimes appeared overly cooperative or ACT-advanced, and scenarios could benefit from greater behavioral inflexibility or richer context. Feedback was occasionally too narrow when therapist utterances reflected preparatory steps rather than complete interventions. Aggregating multiple turns for feedback could better capture broader therapeutic intent.

\subsection{Evaluation of ACT Feedback Model}

We evaluated six LLMs to select an automated evaluator for providing ACT-aligned feedback during interaction. Using 49 psychotherapy transcripts previously rated by a human ACT supervisor \cite{tahir_thinking_2025}, each model independently produced ratings on the Acceptance and Commitment Therapy Fidelity Measure (ACT-FM) and the Therapy Empathy Scale (TES). ACT-FM and TES ratings were computed as described in Section~\ref{rel-work:act}. Model agreement with human ratings was assessed using mean absolute error (MAE), which measures the average absolute difference between model-generated and human ratings (lower indicates closer agreement), and Spearman's $\rho$, which reflects rank-order preservation independent of absolute calibration.

\subsubsection{Full-transcript agreement with human ratings}
Table~\ref{tab:model_compare} summarizes model performance on full-session transcripts. GPT-4o mini achieved the lowest ACT balance MAE (6.12), indicating closest absolute alignment with human supervisor ratings. All models showed statistically significant rank-order correlation with human ratings ($p < .001$). While some models exhibited comparable or higher rank-order agreement (e.g., Grok 4: $\rho = 0.68$), they showed substantially larger MAE values (Grok-4: 9.23), suggesting systematic over- or under-estimation despite preserving relative ordering. GPT-4o-mini also achieved competitive TES MAE (2.28), and its superior ACT-FM calibration makes it the best choice for feedback generation, where ACT fidelity is the primary criterion.

\begin{table}[t]
\centering
\caption{Agreement Between LLM-Generated ACT Fidelity Ratings and Human Supervisor Ratings Across 49 Full Psychotherapy Transcripts. Lower MAE Indicates Closer Calibrated Agreement; Higher $\rho$ Indicates Better Rank-Order Preservation (all $p < .001$)}
\label{tab:model_compare}
\setlength{\tabcolsep}{3pt}
\begin{tabular}{lccc}
\hline
\textbf{Model} 
& \textbf{ACT Balance MAE} $\downarrow$ 
& \textbf{TES MAE} $\downarrow$ 
& $\boldsymbol{\rho} \uparrow$ 
\\ \hline
GPT-4o mini
& \textbf{6.12} 
& 2.28 
& 0.57 \\

GPT-4o 
& 7.78 
& 2.51 
& 0.61 \\

GPT-5.1
& 24.12 
& 2.63 
& 0.55 \\

GPT-5.2 
& 21.68 
& 2.47 
& 0.58  \\

Claude Sonnet 4.5
& 26.01 
& 2.31 
& 0.53 \\

Grok 4
& 9.23 
& \textbf{2.10} 
& \textbf{0.68} \\ \hline
\end{tabular}
\end{table}

\subsubsection{Turn-level and partial-transcript evaluation}
Because the system provides feedback incrementally during live interaction, we evaluated models at the turn level by rating partial transcripts consisting of the first $t$ therapist turns, computing MAE between the model's partial-transcript rating and the human supervisor's full-session rating.

\begin{figure}
    \centering
    \includegraphics[width=\columnwidth]{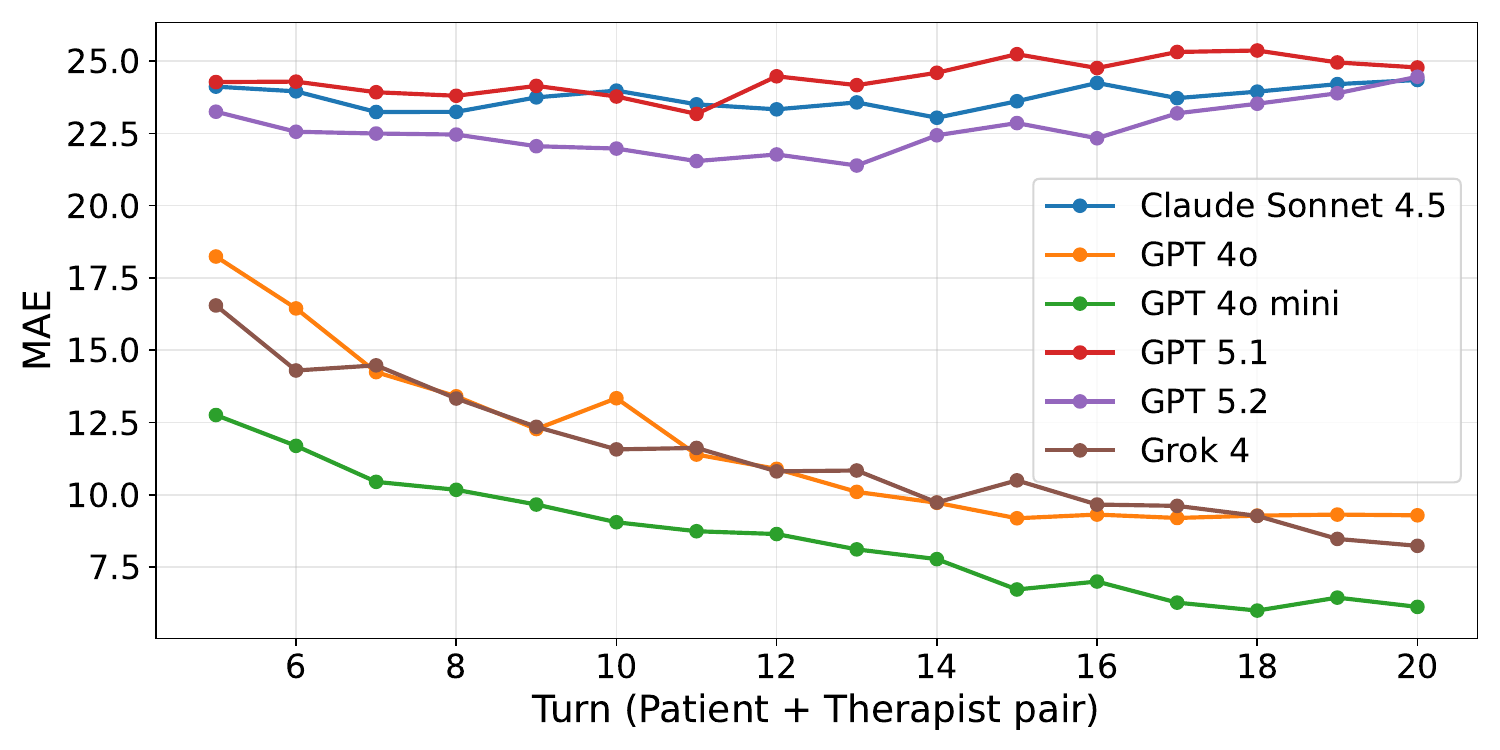}
    \caption{LLM agreement with human supervisor ratings on partial transcripts. Each point shows MAE between the model's ACT balance rating for the first $t$ therapist turns and the human supervisor's full-session rating. Lower values indicate better alignment.}
    \label{fig:prefix_human_mae}
\end{figure}

Models exhibited distinct convergence behaviors as conversational context accumulated (Figure~\ref{fig:prefix_human_mae}). GPT-4o mini showed progressive alignment, with MAE decreasing from 12.76 at turn 5 to 6.12 at turn 20 (6.64-point improvement). GPT-4o and Grok 4 demonstrated similar convergence patterns (8.95 and 8.32-point improvements, respectively). In contrast, larger models (GPT-5.1, GPT-5.2, Claude Sonnet 4.5) showed increasing or stable error with additional context (0.23–1.21-point increases), suggesting sensitivity to early-session content not representative of overall fidelity. Based on its lowest full-session MAE and progressive turn-level convergence, GPT-4o mini demonstrated optimal performance for automated ACT fidelity assessment.

\subsubsection{Latency-accuracy trade-off}
\begin{figure}
    \centering
    \includegraphics[width=\columnwidth]{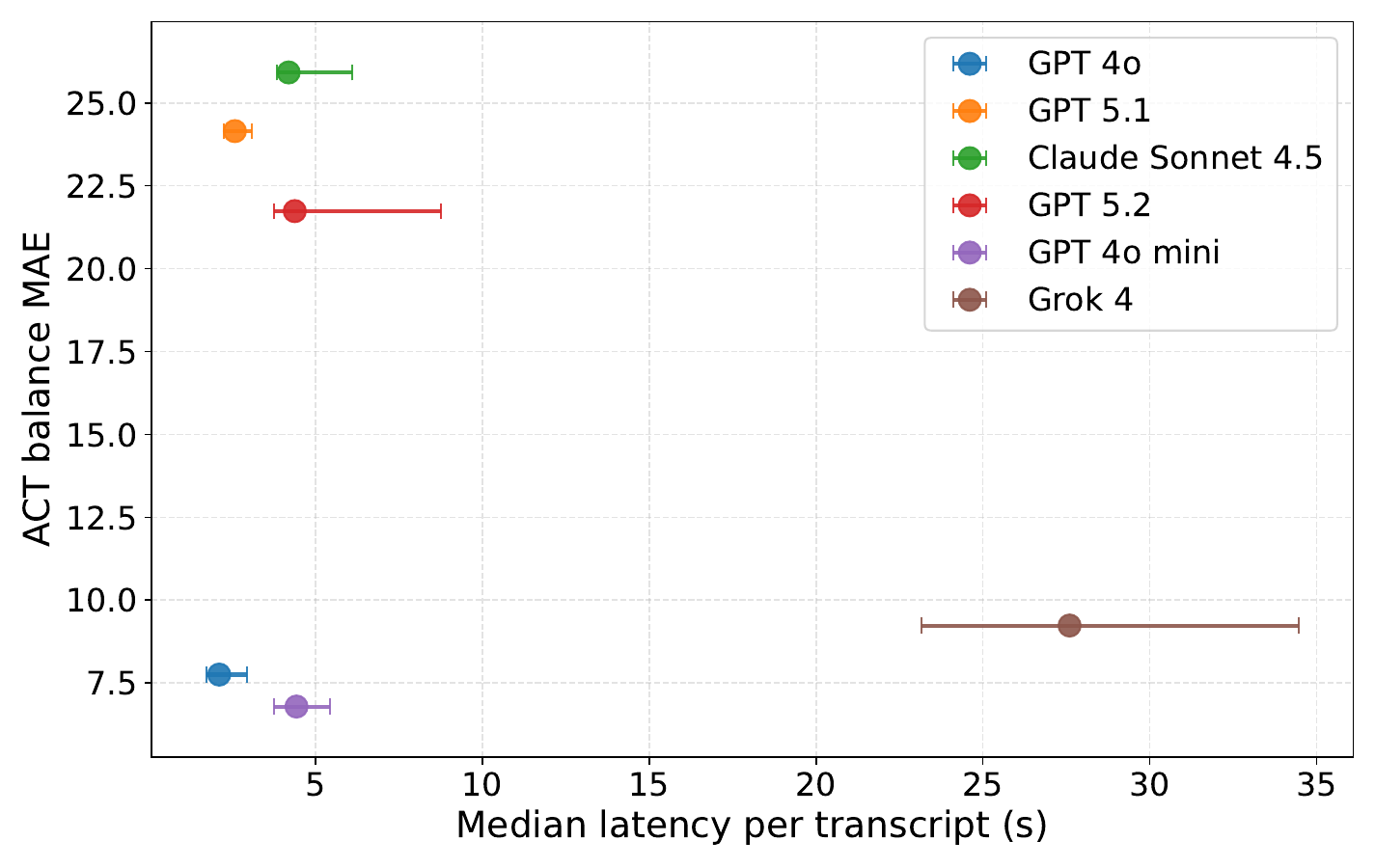}
    \caption{Trade-off between inference speed and rating accuracy for six LLMs. Median latency per transcript is plotted against ACT balance MAE (lower is better). Error bars show P10-P90 latency range.}
    \label{fig:latency_act_mae}
\end{figure}

Inference latency characterizes computational cost and deployment constraints. GPT-4o achieved the lowest median latency (2.1s per transcript, P10-P90: 1.7–2.9s), followed by GPT-5.1 (2.6s, P10-P90: 2.3–3.1s). GPT-4o mini exhibited moderate latency (4.4s, P10-P90: 3.8–5.4s), comparable to Claude Sonnet 4.5 (4.2s, P10-P90: 3.8–6.1s) and GPT-5.2 (4.4s, P10-P90: 3.7–8.8s). Grok 4 showed substantially higher latency (27.6s, P10-P90: 23.2–34.5s). Figure~\ref{fig:latency_act_mae} illustrates this trade-off: while GPT-4o achieved faster inference, it showed higher MAE (7.78) than GPT-4o mini (6.12). In the context of therapist training, we prioritized fidelity-aligned feedback over minimal response time, as feedback is generated after therapist turns and can tolerate modest delays for more reliable evaluation.

\section{Conclusion}

This paper presented an AI-powered simulated patient system for ACT psychotherapy training combining embodied virtual patients with automated real-time fidelity feedback. Expert evaluation with practicing psychologists confirmed realistic patient interactions and effective turn-by-turn feedback that increases therapeutic awareness and supports deliberate practice. Quantitative evaluation across 49 therapy sessions identified GPT-4o mini as achieving closest alignment with human supervisor ACT fidelity ratings (MAE = 6.12, $p < 0.001$). Limitations include occasional overly cooperative patient behavior and narrow feedback for exploratory utterances. Future work should expand patient profile diversity, implement adaptive trajectories, and conduct larger-scale user studies. These findings demonstrate the feasibility of fidelity-aware, AI-supported psychotherapy training as a scalable complement to traditional supervision.

\section*{Acknowledgment}

We thank the participating psychologists for their time, expertise, and thoughtful feedback during the evaluation of the system.


\end{document}